\documentclass[12pt,oneside]{article}
\usepackage{epsfig, cite}
\usepackage{color}
\usepackage{ifthen}
\usepackage{graphicx}
\usepackage{amssymb}

\def\half{{1\over 2}}
\def\({\left(}
\def\){\right)}
\def\[{\left[}
\def\]{\right]}

\def\e{\begin{equation}}
\def\q{\end{equation}}
\def\m{\begin{eqnarray}}
\def\n{\end{eqnarray}}

\begin{document}
\thispagestyle{empty} \setcounter{page}{0}
\renewcommand{\theequation}{\thesection.\arabic{equation}}

\vspace{2cm}

\begin{center}
{\huge The Curvature Perturbation in the Axion-type Curvaton Model}

\vspace{1.4cm}

Pravabati Chingangbam and Qing-Guo Huang

\vspace{.2cm}

{\em School of physics, Korea Institute for
Advanced Study,} \\
{\em 207-43, Cheongryangri-2 Dong,
Dongdaemun-Gu, } \\
{\em Seoul 130-722, Korea}\\
\end{center}

\vspace{-.1cm}

\centerline{{\tt prava@kias.re.kr}} \centerline{{\tt
huangqg@kias.re.kr}} \vspace{1cm} \centerline{ABSTRACT}
\begin{quote}
\vspace{.5cm}

We study the axion-type curvaton model, with emphasis on the large
field regime where analytic results are very difficult to obtain. We
evaluate the tensor-scalar ratio $r$ using WMAP normalization and
the non-linearity parameters $f_{NL}$ and $g_{NL}$ by solving the
equations numerically using the $\delta N$ formalism. We compare
them with results for the curvaton with quadratic potential. We find
that $r$ is much smaller for the axion-type case compared to the
result from the quadratic potential, with the difference
increasingly more pronounced at larger field values. $g_{NL}$ is
found to be positive, unlike the quadratic case where it is
negative, and the amplitude of $g_{NL}$ is much larger.  Moreover,
there is a nearly linear scaling between $g_{NL}$ and $f_{NL}$, with
small deviation from linearity at large field values. The slope
between $g_{NL}$ and $f_{NL}$ depends on the parameters
characterizing the axion-type curvaton model. We further consider a
mixed scenario where both the inflaton and the curvaton contribute
to the primordial power spectrum and the non-Gaussianity parameters
are found to be much larger than those in the case with quadratic
potential.

\end{quote}
\baselineskip18pt

\noindent

\vspace{5mm}

\newpage

\setcounter{equation}{0}
\section{Introduction}

Despite the simplicity of the inflationary paradigm \cite{Guth:1980zm},
the mechanism
by which curvature perturbations are generated is not yet fully
established. An alternative to the standard scenario where the
inflaton field is responsible for the accelerated expansion of the
universe as well as the generation of curvature perturbations, is the
curvaton mechanism \cite{Linde:1996gt}. This scenario has been gaining
increasing popularity recently.
According to this scenario the final perturbations are produced from
initial isocurvature perturbation associated with quantum
fluctuations of a light scalar field, the so-called curvaton, whose
energy density is negligible during inflation. These curvaton
isocurvature perturbations are transformed into adiabatic ones when
the curvaton decays into radiation long after the end of inflation.
It is also conceivable that the final curvature perturbations arise
from a mixture of inflaton and curvaton perturbations.
The curvaton scenario may naturally lead to much higher levels of
local-type non-Gaussianity than standard single-field inflation.

The local-type non-Gaussianity can be characterized by some
non-linearity parameters, namely, $f_{NL}$, $g_{NL}$ and so on. They
appear in the  non-linear expansion of the curvature perturbation as:
\e \zeta({\bf x})=\zeta_g({\bf x})+{3\over
5}f_{NL}\zeta_g^2({\bf x})+{9\over 25}g_{NL}\zeta_g^3({\bf x})+...,
\q where $\zeta$ is the curvature perturbation in the uniform
density slice, $\zeta_g$ denotes the Gaussian part of $\zeta$. The
primordial power spectrum, bispectrum and trispectrum are defined by
\m \langle\zeta({\bf k_1})\zeta({\bf k_2})\rangle&=&(2\pi)^3 {\cal
P}_{\zeta}(k_1)\delta^3({\bf k_1}+{\bf k_2}), \\
\langle\zeta({\bf
k_1})\zeta({\bf k_2})\zeta({\bf k_3})\rangle&=&(2\pi)^3
B_\zeta(k_1,k_2,k_3)\delta^3({\bf k_1}+{\bf k_2}+{\bf k_3}), \\
\langle\zeta({\bf k_1})\zeta({\bf k_2})\zeta({\bf k_3})\zeta({\bf
k_4})\rangle&=&(2\pi)^3 T_\zeta(k_1,k_2,k_3,k_4)\delta^3({\bf
k_1}+{\bf k_2}+{\bf k_3}+{\bf k_4}).
\n
The bispectrum and
trispectrum are respectively related to the power spectrum by
\m
B_\zeta(k_1,k_2,k_3)&=&{6\over 5}
f_{NL}[{\cal P}_\zeta(k_1){\cal P}_\zeta(k_2)+2\ \hbox{perms}], \\
T_\zeta(k_1,k_2,k_3,k_4)&=&\tau_{NL}[{\cal P}_\zeta(k_{13}){\cal
P}_\zeta(k_3){\cal P}_\zeta(k_4)+11\ \hbox{perms}] \nonumber \\
&+&{54\over 25}g_{NL}[{\cal P}_\zeta(k_2){\cal P}_\zeta(k_3){\cal
P}_\zeta(k_4)+3\ \hbox{perms}].
\n
The parameter $\tau_{NL}$ depends on $f_{NL}$, as
 \e \tau_{NL}={36\over 25}f_{NL}^2, \q
if the total curvature perturbation is generated by only one field.
The parameters $f_{NL}$, $\tau_{NL}$ and $g_{NL}$ parameterize the
non-Gaussianty from the irreducible three-point and four-point
correlation functions respectively. See \cite{Bartolo:2004if} for
more discussions on the bispectrum and trispectrum.

In the simplest version of inflation model $f_{NL}\sim {\cal
O}(n_s-1)$ \cite{Maldacena:2002vr}. It is constrained by WMAP5
$(n_s=0.96_{-0.013}^{+0.014})$ \cite{Komatsu:2008hk} to be much less
than unity.  Even though a Gaussian distribution of the primordial
curvature perturbation is still consistent with WMAP5
$(-9<f_{NL}<111, \ \hbox{at} \ 95\% \ \hbox{C.L.})$,  the limit on
$f_{NL}$ on the negative side has been greatly reduced from WMAP3 to
WMAP5, while still allowing for large positive value. The latest
limit on this parameter is $f_{NL}=38\pm 21$ at $68\%$ CL for the
local shape non-Gaussianity in \cite{Smith:2009jr}. This has
prompted many authors to consider different possible mechanisms for
generating a large positive non-Gaussianity
\cite{Huang:2008ze,Huang:2008rj,Enqvist:2008gk,Huang:2008bg,Huang:2008zj,
Kawasaki:2008mc,Langlois:2008vk }.

The curvaton is a light scalar field whose mass is small compared to
the Hubble parameter $H_*$ during inflation. It is sub-dominant
during inflation and its fluctuations are initially of isocurvature
type. After the end of inflation it is supposed to completely decay
into thermalized radiation before primordial nucleosynthesis and the
isocurvature perturbations generated by the curvaton are converted
into final adiabatic perturbations. In order to prevent large
quantum corrections to the curvaton mass and keep it small enough
during inflation some symmetries are called for. It seems natural to
invoke supersymmetry. However, supersymmetry must be broken above
the inflation scale and the mass square of each scalar field
generically receives a correction of order $H_*^2$. So supersymmetry
cannot be invoked to protect the smallness of curvaton mass.

A promising curvaton candidate is the
pseudo-Nambu-Goldstone boson -- axion \cite{Dimopoulos:2003az}, whose
potential for the canonical field $\sigma$ takes the form
\e
V(\sigma)=m^2f^2(1-\cos \frac{\sigma}{f}), \label{axpt}
\q
where $f$ is called the axion decay constant.
Actually the axion field is generic in various string models
\cite{Svrcek:2006yi}. The smallness of the axion mass is protected
by the shift symmetry $\sigma\rightarrow \sigma+\delta$. If
$\sigma\ll f$, the curvaton potential is approximately quadratic,
and the second order non-linearity parameter $g_{NL}$ is expected to be
small. However, $g_{NL}$ is large if the non-quadratic term
in the curvaton potential plays a significant role
\cite{Enqvist:2008gk, Huang:2008bg,Huang:2008zj}.

In the axion-type curvaton model, it is very hard to get analytic
results if $\sigma\sim f$. This regime of large field value was
studied numerically in \cite{Kawasaki:2008mc} where the authors
calculated $f_{NL}$. They concluded that it is quite similar to that
of the quadratic potential if the curvaton accounts for the observed
curvature perturbations. In this paper the curvature perturbation in
the axion-type curvaton model will be studied in more detail, and we
will see that the WMAP normalization and the second order
non-linearity parameter $g_{NL}$ are quite different from those of
the quadratic one, in the large field regime. The curvature
perturbation is calculated numerically using the $\delta N$
formalism~\cite{Starobinsky:1986fxa} and the results are compared
with the quadratic model. We find that, as expected, $r$ agrees with
the quadratic case at small field values. However, it is much
smaller for larger field values and the difference increases rapidly
as the field value increases. $g_{NL}$ is positive and its amplitude
is much larger than the quadratic case. We find that $g_{NL}$ scales
linearly with $f_{NL}$, with small deviation from linearity at large
field values. The linear behavior is similar to the quadratic
potential but the slope is quite different. We also analyze a mixed
scenario where the primordial power spectrum has contributions from
both the inflaton and the curvaton. We find that $g_{NL}$ and
$f_{NL}$ are greatly suppressed with respect to the case where only
the curvaton contributes. However, the suppression is much less for
axion compared to the quadratic potential.

This paper is organized as follows: in Sec. 2, some analytic results
for the curvaton model will be briefly reviewed. In Sec. 3, we
numerically calculate the curvature perturbation for the axion-type
curvaton model based on the $\delta N$ formalism
\cite{Starobinsky:1986fxa}. We calculate $f_{NL}$ and $g_{NL}$ and
demonstrate the results in comparison to the quadratic potential and
the potential with quartic correction. Next, we discuss the mixed
scenario where the total primordial power spectrum has contributions
from both the inflaton as well as the curvaton. Our results are
summarized in Sec. 4. In an appendix we present a general discussion
for the curvaton model with a small correction to the quadratic
potential.

\setcounter{equation}{0}
\section{Analytic results for the curvaton model}

We begin with a brief review of the curvaton model with quadratic
potential. In the limit $\sigma\ll f$ the axion-type curvaton
potential is reduced to this case. The amplitude of the primordial
power spectrum generated by the curvaton with quadratic potential \e
V(\sigma)=\half m^2\sigma^2 \q is given by \e
P_{\zeta_\sigma}={1\over 9\pi^2}f_D^2 {H_*^2\over \sigma_*^2}, \q
where $H_*$ and $\sigma_*$ are the Hubble parameter and the vacuum
expectation value of curvaton field evaluated at Hubble exit during
inflation, respectively, and \e \label{dfd}
f_D={3\Omega_{\sigma,D}\over 4-\Omega_{\sigma,D}}, \quad
\Omega_{\sigma,D}={\rho_\sigma\over \rho_{tot}} \q evaluated at the
time of curvaton decay. WMAP normalization \cite{Komatsu:2008hk}
requires \e P_{\zeta,obs}=2.457_{-0.093}^{+0.092}\times 10^{-9}. \q
The amplitude of the tensor perturbation only depends on the scale
of inflation, as, \e P_T={H_*^2/M_p^2\over \pi^2/2}. \q The
tensor-scalar ratio is defined as $r\equiv P_T/P_\zeta$ and then the
WMAP normalization $(P_{\zeta}=P_{\zeta,obs})$ implies that the
Hubble scale during inflation is related to $r$ by \e H_*\simeq
1.1\times 10^{-4}r^\half M_p. \q Usually the dimensionless parameter
$r$ is used to characterize the inflation scale $H_*$. If the total
amplitude of the primordial power spectrum is generated by the
curvaton, namely $P_{\zeta_\sigma}=P_{\zeta,obs}$, then $r$ is given
by \e r={18\over f_D^2}{\sigma_*^2\over M_p^2}, \label{rfs} \q which
encodes the WMAP normalization. Since the potential takes the
quadratic form, the curvaton field evolves linearly after inflation
and the non-Gaussianity parameters $f_{NL}$ and $g_{NL}$ are
respectively given by \cite{Sasaki:2006kq}, \m
f_{NL}&=&{5\over 4f_D}-{5\over 3}-{5f_D\over 6}, \\
g_{NL}&=&{25\over 54}\(-{9\over f_D}+\half+10f_D+3f_D^2\).
\label{qdg}\n For large positive $f_{NL}$, $f_D\ll 1$, which implies
\e g_{NL}\simeq -{10\over 3}f_{NL}. \q Since $g_{NL}$ is the
coefficient of $\zeta_{g}^3$, it is very difficult to measure
$g_{NL}$ in the near future if it is the same order of $f_{NL}$.

After inflation the universe is dominated by radiation and hence the
Hubble parameter goes like $a^{-2}$. Roughly, the curvaton does not
evolve until $H=m$. Once the Hubble parameter drops below the
curvaton mass, the curvaton field rolls down its potential and
starts to oscillate around its minimum. The energy density of an
oscillating curvaton field with quadratic potential goes like
dust-like matter $(\sim a^{-3})$ and grows with respective to that
of radiation. When the Hubble parameter is roughly the same as the
curvaton decay rate $\Gamma_\sigma$, the curvaton starts to decay
into radiation. For simplicity, the curvaton is assumed to suddenly
decay into radiation at the time of $H=\Gamma_\sigma$. Let us assume
the scale factor equals to 1 at the time of $H=m$, and the scale
factor is denoted by $a$ when $H=\Gamma_\sigma$.  If $\sigma_*\ll
M_p$, the radiation and curvaton energy densities are respectively
$3M_p^2m^2$ and $\half m^2\sigma_*^2$ at the time of $H=m$, and
thus, \e 3M_p^2m^2a^{-4}+\half
m^2\sigma_*^2a^{-3}=3M_p^2\Gamma_\sigma^2. \q The curvaton energy
density parameter at the time of its decay can be written by \e
\Omega_{\sigma,D}={m^2\sigma_*^2\over 6M_p^2\Gamma_\sigma^2}a^{-3}.
\q If  ${\Gamma_\sigma\over m}\ll {\sigma_*^4\over M_p^4}$, then,
$a\simeq ({m^2\sigma_*^2\over 6M_p^2\Gamma_\sigma^2})^{1/3}$ and
$\Omega_{\sigma,D}\simeq 1$, which gives, $f_{NL}\simeq -5/6$. If,
on the other hand, ${\Gamma_\sigma\over m}\gg {\sigma_*^4\over
M_p^4}$, then, $a=\sqrt{m/\Gamma_\sigma}$, and \e
\Omega_{\sigma,D}\simeq {\sigma_*^2\over 6M_p^2}\sqrt{m\over
\Gamma_\sigma}. \q This gives \e f_D\simeq {3\over
4}\Omega_{\sigma,D}\simeq {\sigma_*^2\over 8M_p^2}\sqrt{m\over
\Gamma_\sigma}, \label{fdt} \q which is much smaller than 1, and
hence we get $f_{NL}\simeq {5\over 4f_D}\gg 1$. In this paper we
focus on the latter case with large positive $f_{NL}$. Combining
Eqs.(\ref{rfs}) and (\ref{fdt}) we get \e r=1152\
{\Gamma_\sigma\over m}{M_p^2\over \sigma_*^2}. \label{rgm}\q
$f_{NL}$ and $g_{NL}$ are respectively given by \m
f_{NL}&\simeq&10\sqrt{\Gamma_\sigma\over m}\ {M_p^2\over \sigma_*^2},\\
g_{NL}&\simeq&-{100\over 3}\sqrt{\Gamma_\sigma\over m}\ {M_p^2\over
\sigma_*^2}. \n We see that WMAP normalization implies \e
{\Gamma_\sigma\over m}\simeq 10^{-4}\({r\over f_{NL}}\)^2. \q Then,
if $r\sim 10^{-2}$ and $f_{NL}\sim 10^2$, $\Gamma_\sigma/m\sim
10^{-12}$.

In order to compare with the results of the axion-type curvaton
model, it is better to re-scale the curvaton field to be
$\theta\equiv \sigma/f$, and we get \m f_D&\simeq& {\theta_*^2\over
8}{f^2\over M_p^2}\sqrt{m\over \Gamma_\sigma},\label{afd}
\\ r&=&{18\over f_D^2}{f^2\over M_p^2}\theta_*^2. \n In this paper, we
choose $f/M_p=5\times 10^{-3}$ and $\Gamma_\sigma/m=10^{-4}$ which
are the same as those in \cite{Kawasaki:2008mc}. This gives,
$f_D=3.125\times 10^{-4}\ \theta_*^2$, and $r=4.6\times 10^3 \
\theta_*^{-2}$. For $\theta_*\ll 1$, we get $r \gtrsim 10^3$, which
is not compatible with the bound from WMAP5 of $r<0.20$. On the
other hand, if $r$ is required to satisfy the bound from WMAP5, the
curvature perturbation generated by curvaton will be much smaller
than  $P_{\zeta,obs}$ for $\theta_*\ll 1$. We will more carefully
discuss this case in Sec. 3.2. If we tune $\Gamma_\sigma/m$ to be
very small, for example $10^{-12}$, $r<0.2$ and
$P_{\zeta_\sigma}=P_{\zeta,obs}$ can be achieved. But numerical
calculation becomes much more difficult as we decrease
$\Gamma_{\sigma}/m$ and is not much more instructive. So, in this
paper, we keep $f/M_p=5\times 10^{-3}$ and $\Gamma_\sigma/m=10^{-4}$
fixed and illustrate the physics.

In \cite{Huang:2008bg}, the cosine-type potential of curvaton is
expanded to the second order,
\e V\simeq \half
m^2f^2\theta^2\(1-{1\over 12}\theta^2+...\),
\q
and the second term in the bracket is taken as a correction for small
$\theta$ values. Taking into account the
non-linear evolution of curvaton after inflation but prior to
oscillation, the non-linearity parameters are given by
\m
f_{NL}&=&{5\over 4f_D}(1+h_2)-{5\over 3}-{5f_D\over 6}, \label{fnlc}\\
g_{NL}&=&{25\over 54}\({9\over 4f_D^2}(h_3+3h_2)-{9\over
f_D}(1+h_2)+\half (1-9h_2)+10f_D+3f_D^2\). \label{gnlcx} \n where \m
h_2&=&{1.135\delta(0.951+0.189 \delta)\over (0.951+0.568\delta)^2},\\
h_3&=&{1.135\delta(0.951+0.189 \delta)^2\over (0.951+0.568\delta)^3},\\
\delta&=&{\theta_*^2\over 24}.
\n
If $\theta_*\ll 1$, the above
results are the same as those of quadratic potential. Since $h_2$ is
positive, $f_{NL}$ is enhanced due to the correction. Another
important feature is that $g_{NL}$ is expected to be large and
positive for large $\theta_*$. Even though this expansion is not
reliable for large values of $\theta$ (see Fig. 1),
\begin{figure}[h]
\begin{center}
\leavevmode \epsfxsize=0.6\columnwidth \epsfbox{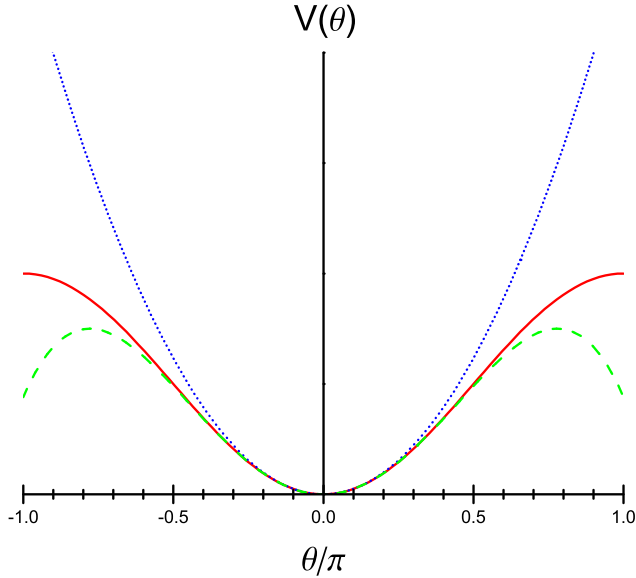}
\end{center}
\caption{The red (solid), blue (dotted) and green (dashed) lines
correspond to the axion potential, quadratic potential and
$\theta^2/2-\theta^4/24$ respectively.}
\end{figure}
(note, in particular, that the potential $\theta^2/2-\theta^4/24$
approaches its maximum value at around $\theta=\pm 0.78\pi$), we will naively
use the above formula to calculate $f_{NL}$ and $g_{NL}$, and
compare them to the numerical results in Sec. 3.1.

Before we end this section, we would like to mention the vacuum
expectation value (VEV) of the curvaton during inflation. At the
classical level, the VEV of the curvaton is taken as a free
parameter. However, in a long-lived inflationary universe, the long
wavelength modes of the quantum fluctuation of a light scalar field
may play a crucial role in its behavior, because its Compton
wavelength is large compared to the Hubble size during inflation. In
\cite{Bunch:1978yq}, the vacuum expectation value of the square of
such a light scalar field with potential $V(\sigma)=\half
m^2\sigma^2$ is given by \e \langle \sigma^2 \rangle={3H_*^4\over
8\pi^2m^2}, \q where $H_*$ is the Hubble parameter during inflation.
Typically the vacuum expectation value of $\sigma$ can be estimated
as \e \sigma_*=\sqrt{\langle \sigma^2\rangle}=\sqrt{3\over
8\pi^2}{H_*^2\over m}. \q For a curvaton field with quadratic
potential, its typical vacuum expectation value is given by
$H_*^2/m$ and its energy density is roughly $H_*^4$. Larger the
Hubble parameter during inflation, larger the energy density of
curvaton. However, the energy density of axion-type curvaton is
bounded by $m^2f^2$ from above. If $\,mf\gg H_*^2$,  $\,\sigma_*$
can be estimated as $H_*^2/m$. On the other hand, $\sigma_*\sim f$
if $mf\leq H_*^2$. Since it is difficult to get analytic results for
the axion-type curvaton model if $\,\sigma_*\sim f$, we will use
numerical method to calculate the curvature perturbation in the next
section.


\setcounter{equation}{0}
\section{Numerical results of the curvature perturbation}

The curvature perturbation on sufficiently large scales on the
uniform density slicing can be computed by the $\delta N$ formalism
\cite{Starobinsky:1986fxa}. Starting from any initial flat slice at
time $t_{ini}$,  on the uniform density slicing, the curvature
perturbation is given by \e \zeta(t, {\bf x})=\delta N=N(t,{\bf
x})-N_0(t), \q where $N(t, {\bf x})=\ln (a(t,{\bf x})/a(t_{ini}))$,
and $N_0(t)=\ln(a(t)/a(t_{ini}))$ is the unperturbed amount of
expansion. In the following subsections we will study two different
curvaton scenarios, first, the curvature perturbation is generated by
the curvaton alone, and secondly, the curvature perturbation is a
mixture of inflaton and curvaton perturbations.

\subsection{Primordial power spectrum mainly generated by curvaton}

In this subsection we focus on the curvaton model in which the
curvature perturbation is generated by a curvaton field. The
curvature perturbation can be expanded as \e
\zeta=N_{,\sigma}\delta\sigma+\half N_{,\sigma\sigma}
{\delta\sigma}^2+{1\over 6}
N_{,\sigma\sigma\sigma}(\delta\sigma)^3+... \, \q where
$N_{,\sigma}=dN/d\sigma$, $N_{,\sigma\sigma}=d^2N/d\sigma^2$ and so
on. Since $\delta\sigma={H_*\over 2\pi}$, the amplitude of the
curvature perturbation generated by curvaton takes the form \e
P_{\zeta_\sigma}=N_{,\sigma}^2\(H_*\over 2\pi\)^2, \q and the
non-Gaussianity parameters are given by \e f_{NL}={5\over
6}{N_{,\sigma\sigma}\over {N_{,\sigma}}^2}, \quad g_{NL}= {25\over
54}{N_{,\sigma\sigma\sigma}\over {N_{,\sigma}}^3}. \q

After inflation, the vacuum energy which governs the dynamics of the
inflaton is converted to radiation and the equations of motion of the
universe are
\m
H^2&=&{1\over 3M_p^2}(\rho_r+\rho_\sigma), \\
\dot \rho_r&+&4H\rho_r=0, \\
\rho_\sigma&=&\half {\dot \sigma}^2+V(\sigma),\\
\ddot\sigma&+&3H\dot \sigma+{dV(\sigma)\over d\sigma}=0, \n where
$\rho_r$ and $\rho_\sigma$ are the energy densities of radiation and
curvaton. It is convenient to define new dimensionless quantities,
namely $N(t)=\ln a(t)$, $x=mt$ and $\theta=\sigma/f$. Then the above
equations of motion are simplified to be \m {dN\over dx}&=&\[\alpha
e^{-4N}+
{f^2\over 3M_p^2}\(\half({d\theta\over dx})^2+{\tilde V}(\theta)\)\]^\half,\\
{d^2\theta\over dx^2}&+&3{dN\over dx}{d\theta\over dx}+{d{\tilde
V}(\theta)\over d\theta}=0, \n where \e V(\sigma)=m^2f^2{\tilde
V}(\theta), \q and \e \alpha={\rho_{r,ini}\over 3M_p^2m^2}, \q and
$\rho_{r,ini}$ is the radiation energy density at $t=t_{ini}$. For
the axion-type curvaton model, ${\tilde V}(\theta)=1-\cos\theta$.
The scale factor can be rescaled to satisfy $a(t_{ini})=1$ and then
$N(t_{ini})=0$. Roughly speaking, the number of e-folds from the end
of inflation to the beginning of the curvaton oscillations is almost
unperturbed and the time when $H=m{dN\over dx}=m$ can be taken as
the initial time $t_{ini}$ for simplicity. Because the energy
density of radiation is still much larger than that of curvaton when
$H=m$, $\rho_{r,ini}=3M_p^2m^2$ and then $\alpha=1$. Here we adopt
the curvaton sudden-decay approximation which says that the curvaton
is proposed to suddenly decay into radiation at the time of
$H=\Gamma_\sigma$. This condition is used to stop the evolution of
curvaton in the numerical calculation. We must mention that all our
numerical results presented in the following are for $f=5\times
10^{-3}M_p$ and $\Gamma_\sigma/m=10^{-4}$, unless otherwise
specified.

\begin{figure}
\begin{center}
\resizebox{280pt}{200pt}{\includegraphics{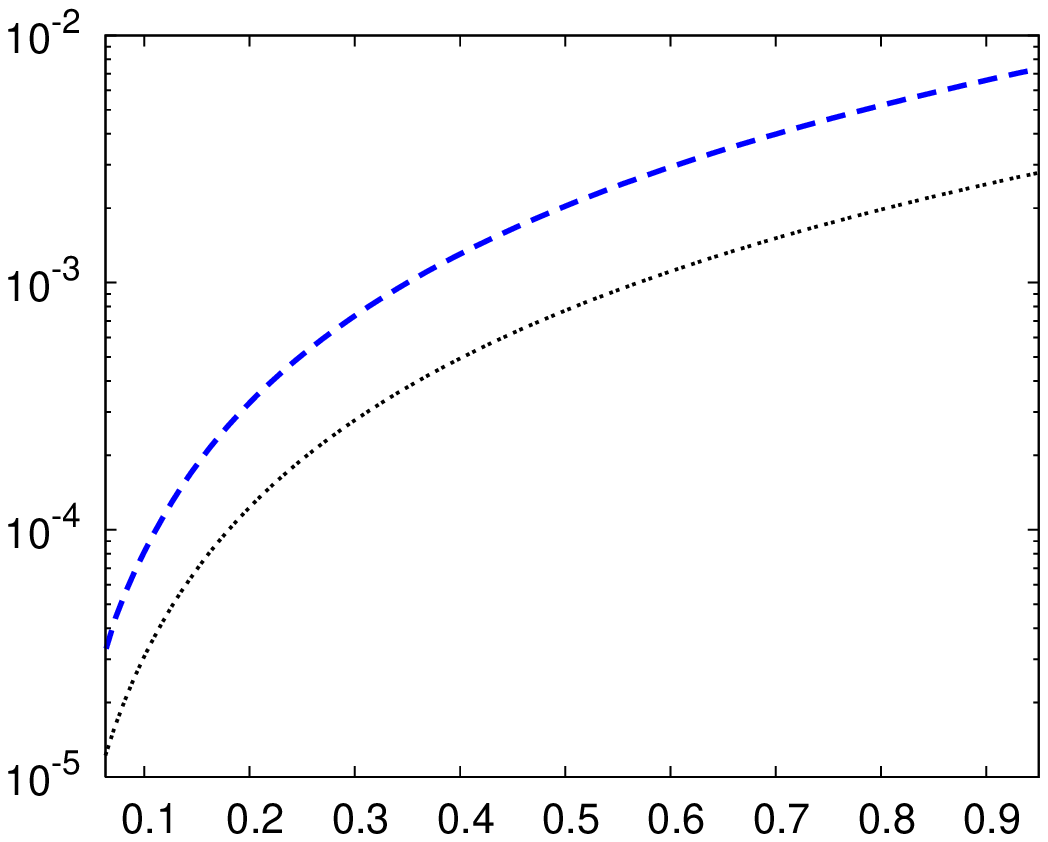}}
\vskip -105 true pt \hskip -320 true pt {\large{$f_D$}} \vskip 95
true pt \hskip 15 true pt {\large{$\theta_*/\pi$}} \vskip 2 true pt
\end{center}
\caption{The blue (dashed) line shows the exact value of
  $f_D$ computed numerically, whereas, the black (dotted) line shows
to $f_D$ computed from the analytic approximation in Eq.(\ref{afd}),
for the quadratic potential. }
\label{fig:fD}
\end{figure}

First of all, we want to compare the analytic results for the
curvaton model with quadratic potential to the numerical results. In
this case, the non-Gaussianity parameters only depend on the
parameter $f_D$. We numerically calculate exact $f_D$ which is given
in Eq.~(\ref{dfd}) and show it along with the analytic approximate
expression for $f_D$ given in Eq.~(\ref{afd}) in Fig.~\ref{fig:fD}.
We see that the exact numerical result (blue dashed line) is larger
than the analytic approximation (black dotted line) by about factor
of two. The reason is that the energy density of the curvaton, as it
oscillates about its minimum potential, does not exactly decrease as
fast as $a^{-3}$, due to the friction term $3H\dot \sigma$, even
though this term is small compared to the driving term $m^2\sigma$.
So the curvaton energy density at the time of its decay is slightly
larger than the analytic approximation.

\begin{figure}
\begin{center}
\resizebox{280pt}{200pt}{\includegraphics{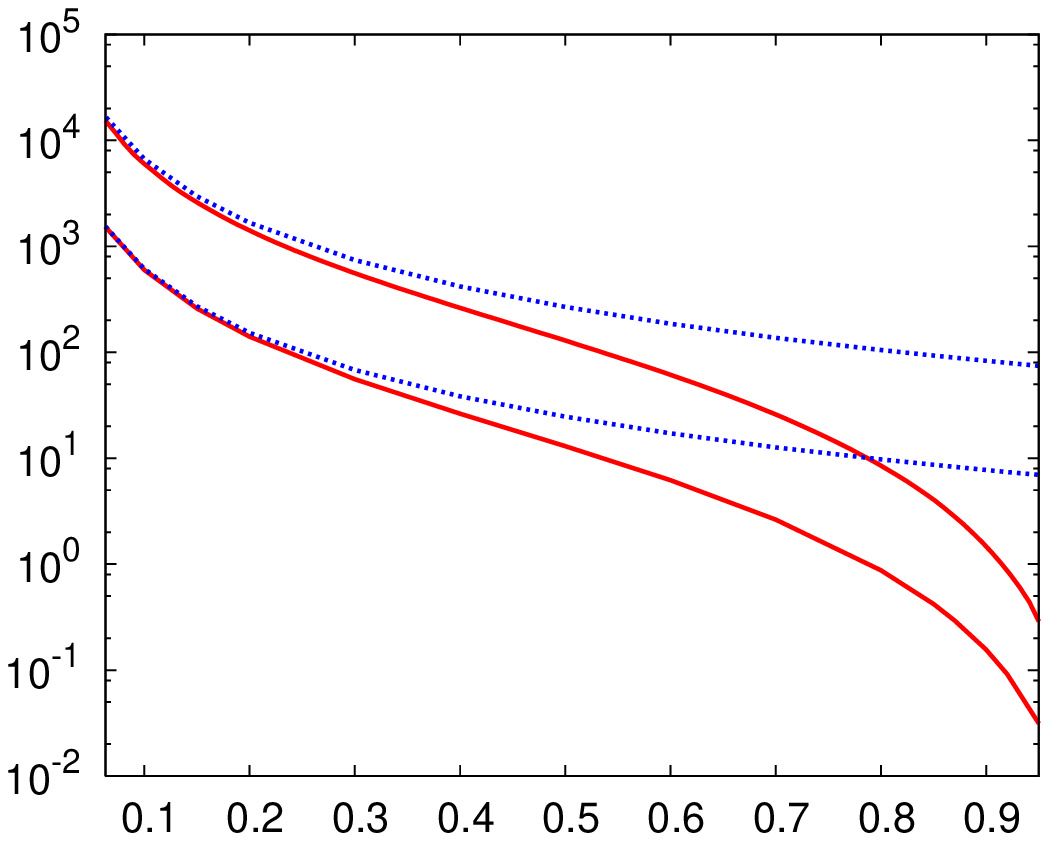}}
\vskip -105 true pt \hskip -320 true pt {\large{$r$}} \vskip 95 true
pt \hskip 15 true pt {\large{$\theta_*/\pi$}} \vskip 2 true pt
\end{center}
\caption{The  red (solid) line shows $r$ for the cosine potential and the
blue (dotted) line for the quadratic potential, respectively, for
$\Gamma_{\sigma}/m = 10^{-4}$ and $10^{-5}$. $r$ is seen to decrease as
$\Gamma_{\sigma}/m $ decreases.}
\label{fig:r}
\end{figure}

Using the dimensionless quantities used in the numerical
calculation, the amplitude of the power spectrum of the curvature
perturbation generated by the curvaton is denoted by \e
P_{\zeta_\sigma}=N_*'^2{H_*^2\over 4\pi^2f^2}, \label{nps} \q where
\e N_*'={dN_{tot}\over d\theta_{ini}}|_{\theta_{ini}=\theta_*}, \q
and $N_{tot}$ denotes the total number of e-folds from $t_{ini}$ to
the time of $H=\Gamma_\sigma$. The tensor-scalar ratio $r$ (or the
inflation scale $H_*$) is determined by WMAP normalization as \e
r={8\over N_*'^2}{f^2\over M_p^2}. \q The numerical result for $r$
is shown in Fig.~\ref{fig:r} where we have plotted for axion-type
(red solid lines) and quadratic case (blue dotted lines). The upper
set of red and blue lines correspond to
 $\Gamma_{\sigma}/m = 10^{-4}$, while the lower set corresponds to
$\Gamma_{\sigma}/m = 10^{-5}$. From this figure we see agreement of
the value of $r$ between the two models for $\theta_*\ll 1$. As
$\theta_*$ increases we find that $r$ for the axion-type potential
increasingly becomes smaller than the quadratic potential. $r$ seems
to scale linearly with $\Gamma_{\sigma}/m$ for the axion-type model,
as is the case for the quadratic potential, as seen in
Eq.~(\ref{rgm}). Hence, for large field values it is possible to
satisfy WMAP normalization and the constraint on $r$ simultaneously
for much larger values of $\Gamma_{\sigma}/m$, as compared to the
quadratic case.

The non-linearity parameters are given by
\e
f_{NL}={5\over 6}{N_*''\over N_*'^2},\quad g_{NL}={25\over 54}
{N_*'''\over N_*'^3}, \q where \m N_*''={d^2N_{tot}\over
d\theta_{ini}^2}|_{\theta_{ini}=\theta_*}, \quad \hbox{and}\quad
N_*'''={d^3N_{tot}\over d\theta_{ini}^3}|_{\theta_{ini}=\theta_*},
\n
\begin{figure}
\begin{center}
\resizebox{280pt}{200pt}{\includegraphics{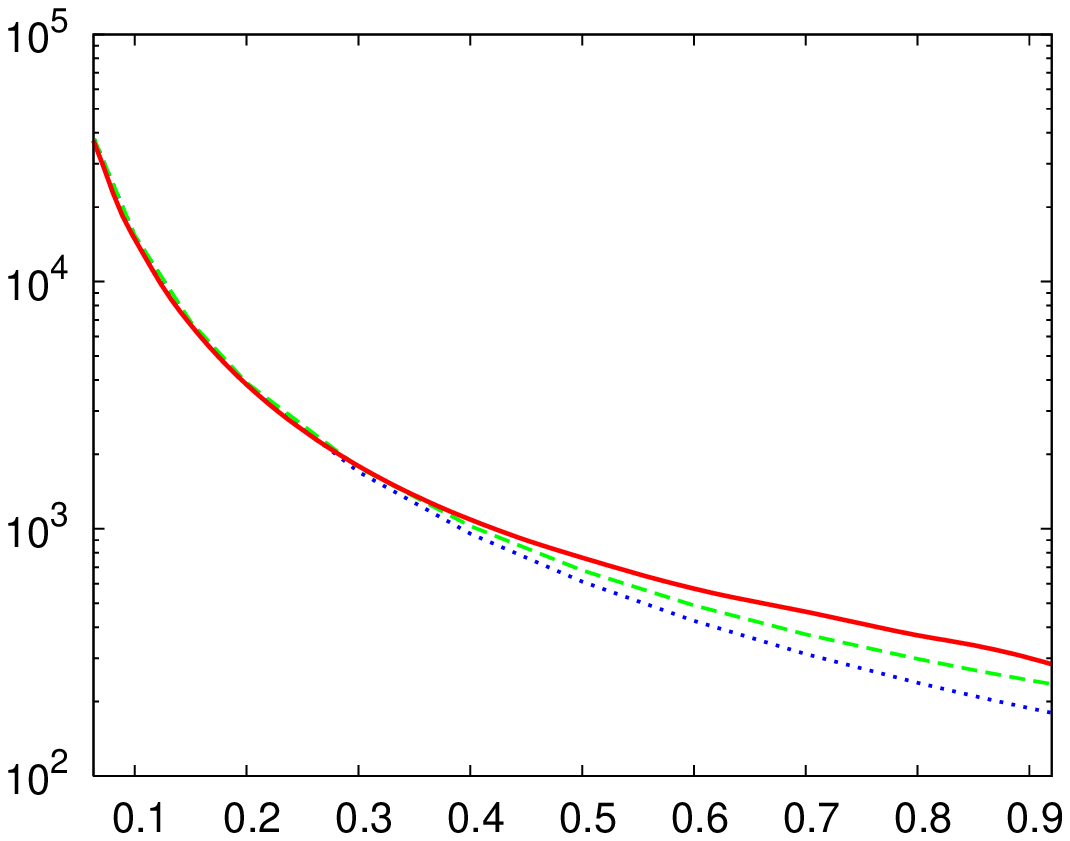}} \vskip
-105 true pt \hskip -320 true pt ${\large{f_{NL}}}$ \vskip 95 true
pt \hskip 15 true pt {\large{$\theta_*/\pi$}} \vskip 2 true pt
\end{center}
\caption{Plot of $f_{NL}$. The red (solid), blue (dotted) and green
 (dashed) lines correspond
to axion-type, quadratic and quadratic with $\sigma^4$ correction models,
respectively.}
\label{fig:fnl}
\end{figure}
Fig.~\ref{fig:fnl} shows the numerical results of $f_{NL}$
 for axion-type (red solid line), quadratic
(blue dotted line), and quadratic with correction (green dashed
line) models. We see that the three results agree very well for
small $\theta_*$ and begins to differ as $\theta_*$ increases. The
difference, however, is only of order one and hence small.
Fig.~\ref{fig:fnl} gives a confirmation of the results
of~\cite{Kawasaki:2008mc}. The correction case is not really correct
at large $\theta_*$ but we still plot it to show the agreement at
small $\theta_*$.

\begin{figure}
\begin{center}
\resizebox{280pt}{200pt}{\includegraphics{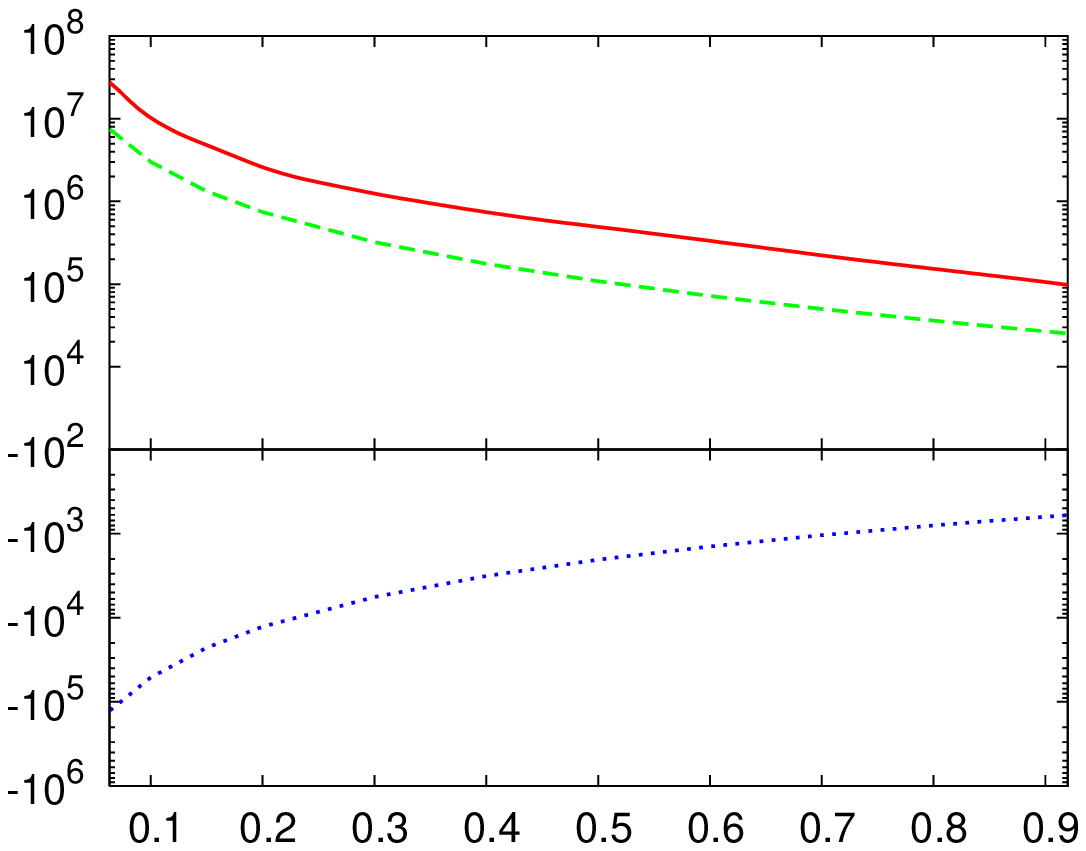}} \vskip
-105 true pt \hskip -320 true pt {\large{$g_{NL}$}} \vskip 95 true
pt \hskip 15 true pt {\large{$\theta_*/\pi$}} \vskip 2 true pt
\end{center}
\caption{Plot of $g_{NL}$. The red (solid), blue (dotted) and green
  (dashed) lines correspond
to axion-type, quadratic and quadratic with  $\sigma^4$ correction
models, respectively. }
\label{fig:gnl}
\end{figure}
Next we look at the results for $g_{NL}$ for the same three models
as above. The numerical results are plotted in Fig.~\ref{fig:gnl}.
The red solid line shows the result for the axion-type curvaton, and
we can see that $g_{NL}$ is positive and larger than the magnitudes
for the other two cases. For the quadratic case, we know from the
analytic expression  of Eq.~(\ref{qdg}) that it is negative and that
is shown by the blue dotted line which gives the exact numerical
answer. Again we know from Eq.~(\ref{gnlcx}) that for correction
case it is positive and that is what we see from the green dashed
line in the figure.

\begin{figure}
\begin{center}
\resizebox{280pt}{200pt}{\includegraphics{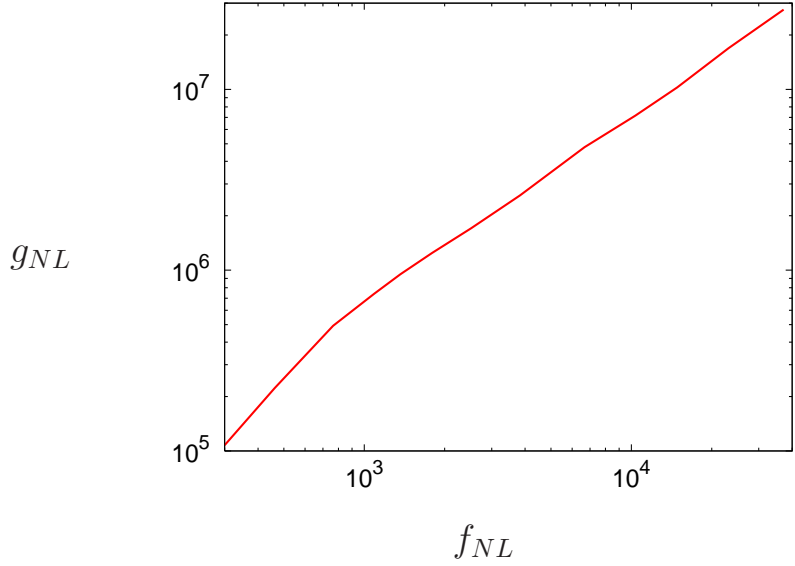}} \vskip -105
true pt \hskip -320 true pt {\large{$g_{NL}$}} \vskip 95 true pt
\hskip 15 true pt {\large{$f_{NL}$}} \vskip 2 true pt
\end{center}
\caption{Plot of $g_{NL}$ $vs.$ $f_{NL}$ for axion-type curvaton. }
\label{fig:fnlgnl}
\end{figure}
From Fig.~\ref{fig:fnlgnl} where we have plotted $g_{NL}$ versus
$f_{NL}$ an interesting observation is that $g_{NL}$ almost linearly
depends on $f_{NL}$ for the axion-type curvaton model, namely \e
g_{NL}\simeq c f_{NL}, \q where $c\simeq 732$. From
Fig.~\ref{fig:fnl} and \ref{fig:gnl}, we can see that the formulas
for the quadratic with correction capture some features of the
axion-type curvaton model. Here we try to use them to roughly
understand why $g_{NL}$ almost linearly depends on $f_{NL}$. At the
leading order, we have \e h_2\simeq h_3\simeq
1.193\delta=0.05\theta_*^2, \q and then \e f_{NL}\simeq
1.25/f_D,\quad g_{NL}\simeq 0.21\theta_*^2/f_D^2. \q Taking into
account the analytic formula for $f_D$ in Eq.~(\ref{afd}), we obtain
\m
f_{NL}&\simeq& 10 {M_p^2\over f^2}\sqrt{\Gamma_\sigma\over m}\theta_*^{-2}, \\
g_{NL}&\simeq&13.4{M_p^4\over f^4}{\Gamma_\sigma\over
  m}\theta_*^{-2}.
\n
Therefore, the coefficient $c$ is roughly given by \e c\simeq 1.34
{M_p^2\over f^2}\sqrt{\Gamma_\sigma\over m}. \q Using $f/M_p=5\times
10^{-3}$ and $\Gamma_\sigma/m=10^{-4}$, $c\simeq 536$ which is
roughly the same as that for the axion-type curvaton model. A
general discussion for the curvaton model with a small correction to
the quadratic potential is given in the Appendix.


\subsection{Mixed scenario}

In the previous subsection, where we assumed that the total
primordial power spectrum is generated by the curvaton only, we saw
that, having fixed $f/M_p=5\times 10^{-3}$, the value of
$\Gamma_\sigma/m=10^{-4}$ is too large for $r$ to satisfy
observational constraints. We know from Eq.~(\ref{rgm}) that $r$
scales linearly with $\Gamma_{\sigma}/m$ for the quadratic case.
From Fig.~\ref{fig:r} it seems that for the axion-type case also $r$
scales linearly with $\Gamma_{\sigma}/m$, though it is not obvious
from the equations. Hence, this value was enough to demonstrate the
physics.

In this subsection we consider the scenario where the final
curvature perturbations are contributed by both inflaton and
curvaton fluctuations. This scenario has been invoked in
\cite{Huang:2008rj} to naturally achieve a red-tilted primordial
power spectrum in the curvaton model. It is convenient to introduce
a new parameter $\beta$, \e \beta={P_{\zeta_\sigma}/ P_\zeta}, \q
where $P_\zeta$ is the total primordial power spectrum including the
contribution from inflaton, and hence $\beta\in [0,1]$. Then the
effective non-Gaussianity parameters becomes \e f_{NL}^{eff}=\beta^2
f_{NL},\quad g_{NL}^{eff}=\beta^3 g_{NL}. \q See
\cite{Huang:2008rj,Huang:2008zj} for detail. The same result of
$f_{NL}^{eff}$ in the mixed scenario is also obtained in
\cite{Langlois:2008vk}. If $g_{NL}=cf_{NL}$, \e g_{NL}^{eff}=\beta
cf_{NL}^{eff} \q where $g_{NL}^{eff}$ is suppressed by a small
factor $\beta$ with respect to $f_{NL}^{eff}$. In
\cite{Huang:2008zj}, the author found \e g_{NL}=cf_{NL}^2 \q for the
curvaton model in which the curvaton has a polynomial potential and
the quadratic term is subdominant during inflation. In this case, \e
g_{NL}^{eff}={c\over\beta}(f_{NL}^{eff})^2, \q which is enhanced by
a factor $1/\beta$. A large $g_{NL}$ is expected in the curvaton
model. In general, $\tau_{NL}^{eff}\geq {36\over
25}(f_{NL}^{eff})^2$ in the multi-field case \cite{Suyama:2007bg}.
In particular, in the mixed scenario, $\tau_{NL}^{eff}$ is enhanced
by a factor $1/\beta$, namely \e \tau_{NL}^{eff}={36\over
25\beta}(f_{NL}^{eff})^2. \q

\begin{figure}
\begin{center}
\resizebox{280pt}{200pt}{\includegraphics{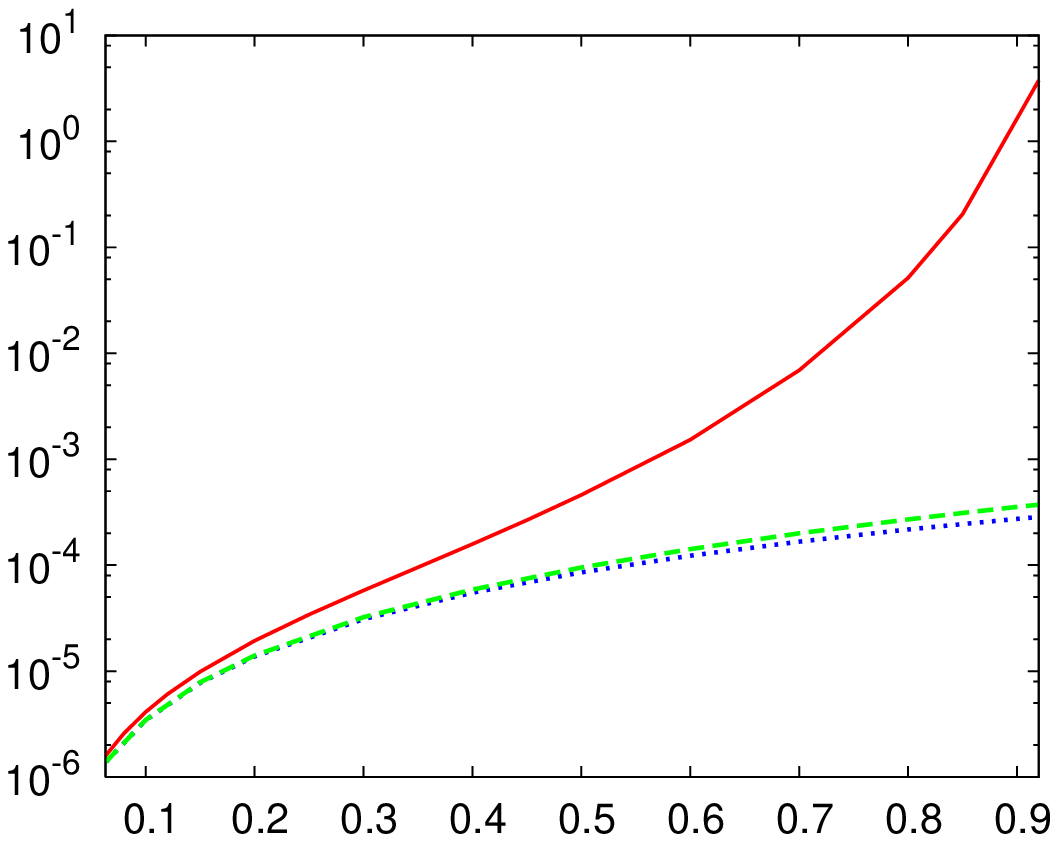}} \vskip
-105 true pt \hskip -320 true pt {\large{$f_{NL}^{eff}$}} \vskip 95 true
pt \hskip 15 true pt {\large{$\theta_*/\pi$}} \vskip 2 true pt
\end{center}
\caption{Plot of $f_{NL}^{eff}$. The red (solid), blue (dotted) and
  green (dashed) lines correspond
to axion-type, quadratic and quadratic with $\sigma^4$ correction models,
  respectively. }
\label{fig:fnleff}
\end{figure}

\begin{figure}
\begin{center}
\resizebox{280pt}{200pt}{\includegraphics{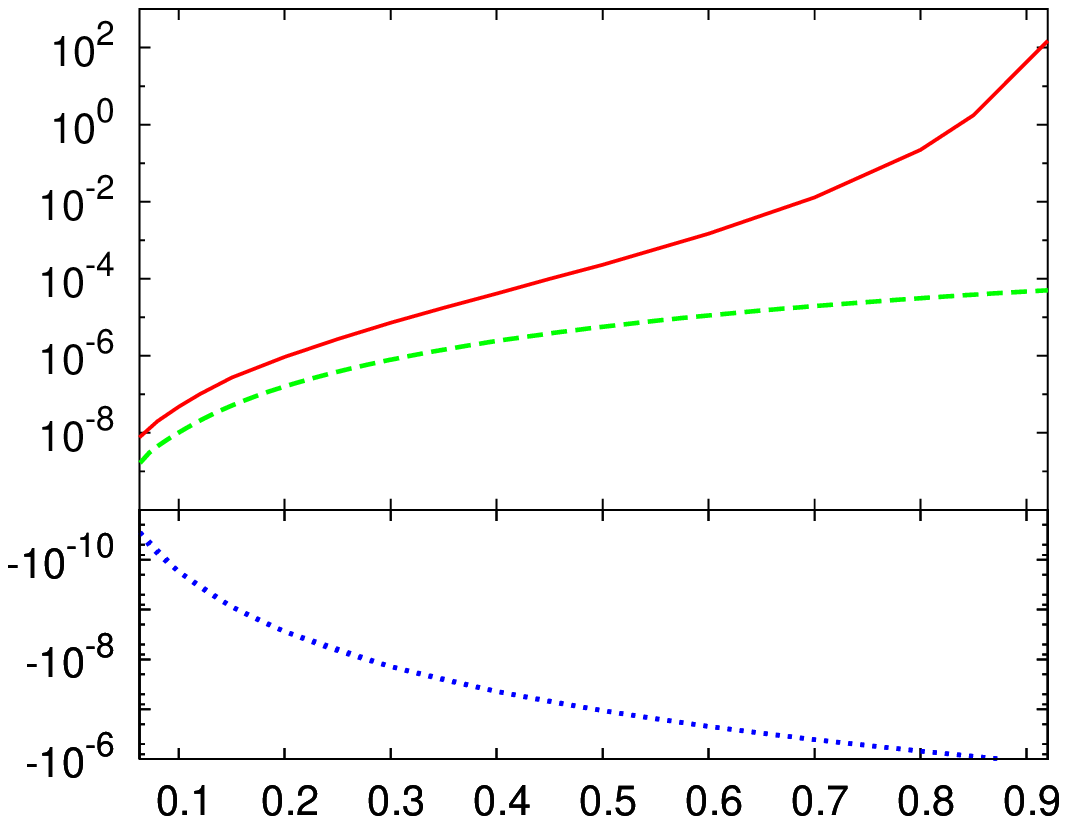}} \vskip
-105 true pt \hskip -320 true pt {\large{$g_{NL}^{eff}$}} \vskip 95 true
pt \hskip 15 true pt {\large{$\theta_*/\pi$}} \vskip 2 true pt
\end{center}
\caption{Plot of $g_{NL}^{eff}$. The red (solid), blue (dotted) and
  green (dashed) lines correspond
to axion-type, quadratic and quadratic with $\sigma^4$ correction models,
  respectively. }
\label{fig:gnleff}
\end{figure}

\begin{figure}
\begin{center}
\resizebox{280pt}{200pt}{\includegraphics{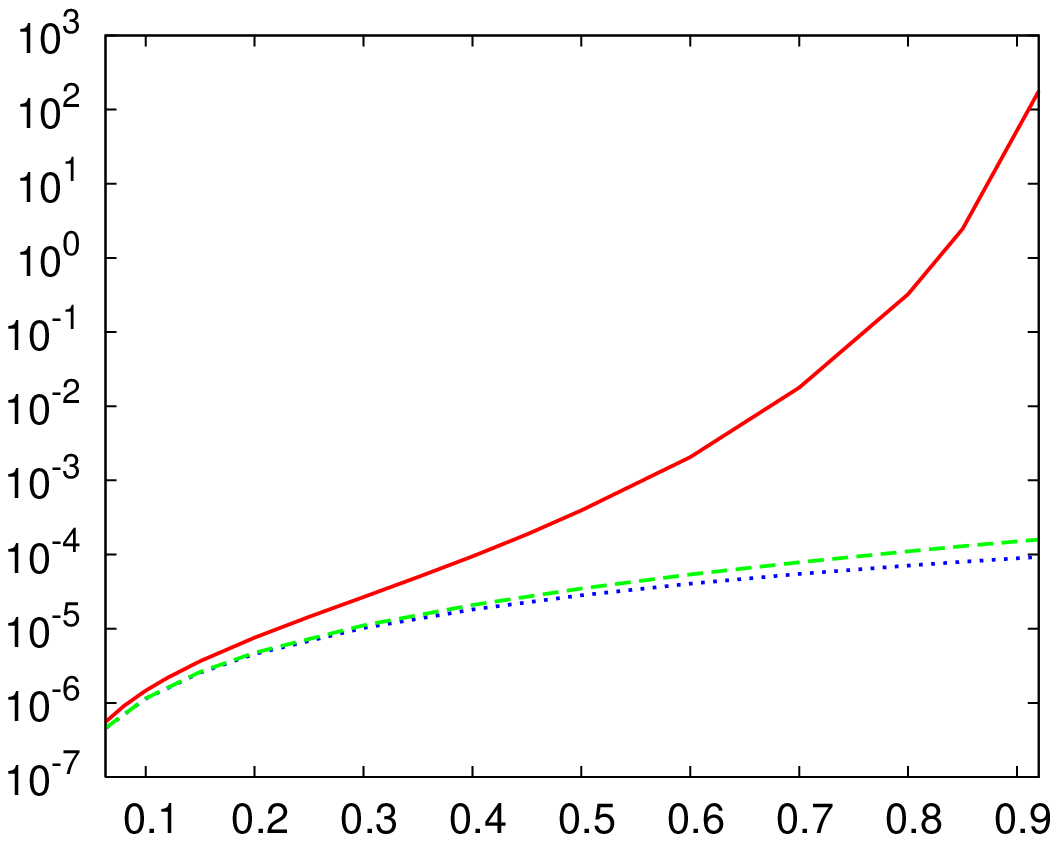}} \vskip -105
true pt \hskip -320 true pt {\large{$\tau_{NL}^{eff}$}} \vskip 95
true pt \hskip 15 true pt {\large{$\theta_*/\pi$}} \vskip 2 true pt
\end{center}
\caption{Plot of $\tau_{NL}^{eff}$. The red (solid), blue (dotted) and
  green (dashed) lines
correspond to axion-type, quadratic and quadratic with $\sigma^4$
correction models,  respectively. }
\label{fig:taunleff}
\end{figure}

For the axion-type curvaton model $\beta$ can be expressed as \e
\beta\equiv{P_{\zeta_\sigma}\over P_\zeta}={r\over
8}N_*'^2{M_p^2\over f^2}. \q If $r<8f^2/(N_*'^2M_p^2)$, then
$\beta<1$, which implies that the quantum fluctuation of the
curvaton only contributes a part of the total primordial curvature
perturbation.
In Figs.~\ref{fig:fnleff}, ~\ref{fig:gnleff} and ~\ref{fig:taunleff},
we have plotted $f_{NL}^{eff}$, $g_{NL}^{eff}$ and $\tau_{NL}^{eff}$,
respectively. The parameter values we have used for these plots are
$f/M_p=5\times 10^{-3}$, $\,\Gamma_\sigma/m=10^{-4}$ and $r=0.1$.
We see that the values are suppressed by
factor of $\beta^2$ for $f_{NL}$ and by $\beta^3$ for $g_{NL}$ and
$\tau_{NL}$.

\setcounter{equation}{0}
\section{Conclusion}

The curvaton model is one of the few models for the generation of
primordial density perturbations which predict large non-Gaussianity
and is potentially testable in the near future in upcoming
experiments such as Planck. Specific models that have been well
studied and for which analytic approximate results can be obtained
are the simplest ones such as a polynomial potential
\cite{Huang:2008zj}. In this paper we solved the evolution equations
for the quadratic potential to see whether the analytic
approximations are really accurate. We found that the analytic
expression underestimates the exact answer by about factor of two.
The reason is that the axion does not exactly behave as dust-like
matter when it oscillates about the minimum of its potential. The
friction term in the axion equation of motion, though small, is not
negligible and makes the energy density decrease with respect to the
scale factor as $a^{-3+\epsilon}$, where $\epsilon$ is small and
positive.

Our main motivation in this paper was to understand the
non-linearity parameter $g_{NL}$ in the axion-type curvaton model.
To achieve this goal we used the $\delta N$ formalism to solve for
the two non-linearity parameters. The axion type model is very hard
to solve analytically for large field values. In addition, computing
$g_{NL}$ numerically is much harder than $f_{NL}$ since it contains
a third order derivative as against the second order derivative of
$f_{NL}$. We have evaluated the tensor-scalar ratio $r$ using WMAP
normalization and used it to compute $f_{NL}$ and $g_{NL}$. We
compare them with results for the curvaton with quadratic potential.
We found that $r$ is much smaller for the axion-type case compared
to the quadratic potential when $\sigma_*\sim f$. The difference
between them becomes increasingly more pronounced at larger field
values. $g_{NL}$ is found to be positive, unlike the quadratic case
where it is negative, and the amplitude of $g_{NL}$ is much larger.
We also found that there is a nearly linear scaling between $g_{NL}$
and $f_{NL}$, with small deviation from linearity at large field
values. Lastly, we considered a mixed scenario where both the
inflaton and the curvaton contribute to the primordial power
spectrum. We have found that $g_{NL}$ and $f_{NL}$ are greatly
suppressed with respect to the case where only the curvaton
contributes. Moreover, $g_{NL}^{eff}$ is suppressed with respect to
$f_{NL}^{eff}$, whereas,  $\tau_{NL}^{eff}$ is enhanced and hence it
would be inconsistent to ignore the contribution to the overall
primordial non-Gaussianity from $\tau_{NL}^{eff}$.

We conclude with a few remarks on the
trispectrum. Usually the trispectrum is characterized by two parameters
$\tau_{NL}$ and $g_{NL}$. While $f_{NL}$ is well studied for a variety of
curvaton models in the literature, the trispectrum is not as well
studied. One reason is the technical difficulty of having to deal
with third order perturbations. Several recent works have shown that
$\tau_{NL}$ and $g_{NL}$ can be as important as, if not more than,
$f_{NL}$. On the observation side much work has been devoted to
constraining the value of $f_{NL}$ using the CMB observations. It
makes sense to ignore $\tau_{NL}$ and $g_{NL}$ if their contribution
is small. However, we need to take them into account if the
trispectrum is large since they are independent parameters and have
the potential to discriminate between different curvaton models. It
would be very interesting to study the implications of large
$\tau_{NL}$ and $g_{NL}$ on the CMBR and we are working along those
lines.

\vspace{.5cm}%

\noindent {\bf Acknowledgments}

We would like to acknowledge use of QUEST, one of the computing
facilities at Korea Institute for Advanced Study, where the
numerical computation in this paper was carried out. QGH would like
to thank K.~Nakayama for useful discussions. We would also like to
thank C.~Byrnes for commenting on the first version of this paper
and pointing out the relevance of $\tau_{NL}$.

\vspace{5mm}

\appendix

\setcounter{equation}{0}
\section{$g_{NL}$ vs. $f_{NL}$ in the curvaton model with small deviation
from quadratic potential}

\begin{figure}
\begin{center}
\resizebox{280pt}{200pt}{\includegraphics{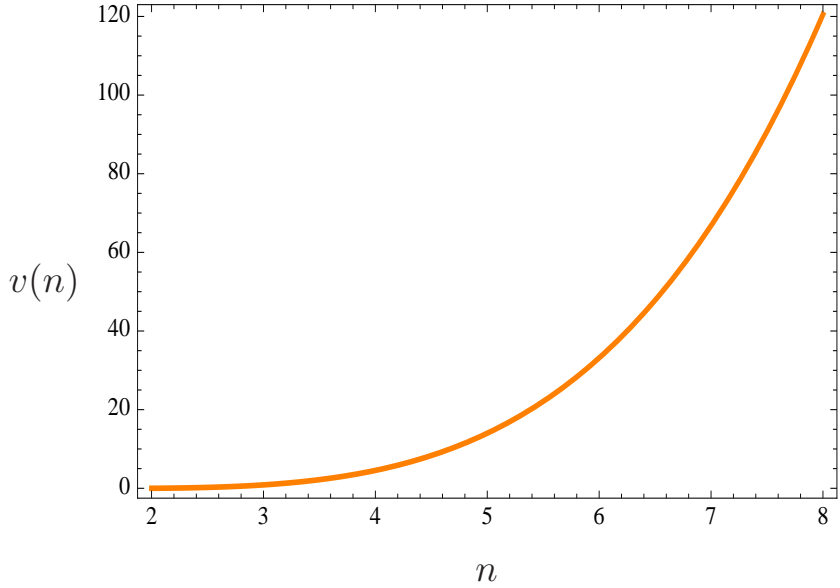}} \vskip -105 true
pt \hskip -320 true pt {\large{$v(n)$}} \vskip 95 true pt \hskip 15
true pt {\large{$n$}} \vskip 2 true pt
\end{center}
\caption{Plot of $v(n)$ as a function of $n$.}
\label{fig:vn}
\end{figure}

The non-Gaussianity for the curvaton model with small deviation from
quadratic potential is investigated in \cite{Huang:2008bg}. In this
appendix we extend the previous discussion. Assume the curvaton
potential is given by \e V(\sigma)=\half m^2\sigma^2+\lambda m^4
\({\sigma\over m}\)^n, \q where $n\geq 2$. If $n<2$, the correction
term becomes dominant around $\sigma=0$. The correction in the
equation of motion is small if $|s|\ll 1/n$, where \e s=\lambda
\({\sigma_*\over m}\)^{n-2}. \q The general result is given in
\cite{Huang:2008bg}. Here $\lambda=-m^2/(24f^2)$ and $s=-\delta$ for
the axion-type curvaton model. For case of the small correction,
$f_{NL}$ and
$g_{NL}$ take the form \m f_{NL}&\simeq& {5\over 4 f_D}, \\
g_{NL}&\simeq&-{25\over 24}v(n) {s\over f_D^2}-{25\over 6f_D}
\label{gnlc} \n where \m v(n)&=&-n^2(n-1)(n-2)
  2^{\frac{n-1}{4}}\Gamma(5/4)^{n-1}\pi \nonumber \\
  &\times& \[J_{\frac{1}{4}}(\half)\int_0^{\half}
J_{\frac{1}{4}}^{n-1}(x)Y_{\frac{1}{4}}(x)x^{\frac{6-n}{4}}dx-
Y_{\frac{1}{4}}(\half)\int_0^{\half}
J_{\frac{1}{4}}^{n}(x)x^{\frac{6-n}{4}}dx \], \n which shows up in
Fig. \ref{fig:vn}.
From Eq.(\ref{gnlc}), $g_{NL}$ becomes positive if \e s<-{4\over
v(n)}f_D. \q If $g_{NL}$ is dominated by the second term in
Eq.(\ref{gnlc}), $g_{NL}\simeq -{10\over 3}f_{NL}$. Now we focus on
the case of $|s|\gg 4f_D/v(n)$, namely \e g_{NL}\simeq -{25\over
24}v(n){s\over f_D^2}. \q Using Eq.(\ref{fdt}), we have \m
f_{NL}&\simeq&10{M_p^2\over m^2}\sqrt{
\Gamma_\sigma \over m}\({\sigma_*\over m}\)^{-2}, \\
g_{NL}&\simeq&-{200\over 3}v(n)\lambda{M_p^4\over m^4}
{\Gamma_\sigma\over m}\({\sigma_*\over m}\)^{n-6}. \n Therefore, \e
g_{NL}\simeq c f_{NL}^{3-n/2}, \q where \e c=-{200\over
3}10^{n/2-3}v(n)\lambda\(M_p\over m\)^{n-2} \({\Gamma_\sigma\over
m}\)^{n-2\over 4}. \q For $n=2$, $c=v(n)=0$ and $g_{NL}$ still
linearly depends on $f_{NL}$. In the mixed scenario, \e
g_{NL}^{eff}=\beta^{n-3} c (f_{NL}^{eff})^{3-n/2}. \q The second
order non-Gaussianity parameter is enhanced in the mixed scenario if
$n<3$. However, for $2\leq n<3$, $v(n)\leq 1$, and hence a large
absolute value of $g_{NL}$ is quite unlikely.



\begin{thebibliography}{99}
\baselineskip=16pt


\bibitem{Guth:1980zm}
  A.~H.~Guth,
  ``The Inflationary Universe: A Possible Solution To The Horizon And Flatness
  Problems,''
  Phys.\ Rev.\  D {\bf 23}, 347 (1981);\\
  A.~D.~Linde,
  ``A New Inflationary Universe Scenario: A Possible Solution Of The Horizon,
  Flatness, Homogeneity, Isotropy And Primordial Monopole Problems,''
  Phys.\ Lett.\  B {\bf 108}, 389 (1982);\\
  A.~Albrecht and P.~J.~Steinhardt,
  ``Cosmology For Grand Unified Theories With Radiatively Induced Symmetry
  Breaking,''
  Phys.\ Rev.\ Lett.\  {\bf 48}, 1220 (1982).



\bibitem{Linde:1996gt}
  A.~D.~Linde and V.~F.~Mukhanov,
  ``Nongaussian isocurvature perturbations from inflation,''
  Phys.\ Rev.\  D {\bf 56}, 535 (1997)
  [arXiv:astro-ph/9610219];\\
  K.~Enqvist and M.~S.~Sloth,
  ``Adiabatic CMB perturbations in pre big bang string cosmology,''
  Nucl.\ Phys.\  B {\bf 626}, 395 (2002)
  [arXiv:hep-ph/0109214];\\
  D.~H.~Lyth and D.~Wands,
  ``Generating the curvature perturbation without an inflaton,''
  Phys.\ Lett.\  B {\bf 524}, 5 (2002)
  [arXiv:hep-ph/0110002];\\
  T.~Moroi and T.~Takahashi,
  ``Effects of cosmological moduli fields on cosmic microwave background,''
  Phys.\ Lett.\  B {\bf 522}, 215 (2001)
  [Erratum-ibid.\  B {\bf 539}, 303 (2002)]
  [arXiv:hep-ph/0110096].


\bibitem{Bartolo:2004if}
  N.~Bartolo, E.~Komatsu, S.~Matarrese and A.~Riotto,
  ``Non-Gaussianity from inflation: Theory and observations,''
  Phys.\ Rept.\  {\bf 402}, 103 (2004)
  [arXiv:astro-ph/0406398];\\
  L.~Alabidi and D.~H.~Lyth,
  ``Inflation models and observation,''
  JCAP {\bf 0605}, 016 (2006)
  [arXiv:astro-ph/0510441];\\
  C.~T.~Byrnes, M.~Sasaki and D.~Wands,
  ``The primordial trispectrum from inflation,''
  Phys.\ Rev.\  D {\bf 74}, 123519 (2006)
  [arXiv:astro-ph/0611075].




\bibitem{Maldacena:2002vr}
  J.~M.~Maldacena,
  ``Non-Gaussian features of primordial fluctuations in single field
  inflationary models,''
  JHEP {\bf 0305}, 013 (2003)
  [arXiv:astro-ph/0210603].


\bibitem{Komatsu:2008hk}
  E.~Komatsu {\it et al.}  [WMAP Collaboration],
  ``Five-Year Wilkinson Microwave Anisotropy Probe (WMAP)
  Observations:Cosmological Interpretation,''
  arXiv:0803.0547 [astro-ph].

\bibitem{Smith:2009jr}
  K.~M.~Smith, L.~Senatore and M.~Zaldarriaga,
  arXiv:0901.2572 [astro-ph].

\bibitem{Huang:2008ze}
  Q.~G.~Huang,
  ``Large Non-Gaussianity Implication for Curvaton Scenario,''
  Phys.\ Lett.\  B {\bf 669}, 260 (2008)
  [arXiv:0801.0467 [hep-th]];\\
  T.~Matsuda,
  ``Modulated Inflation,''
  Phys.\ Lett.\  B {\bf 665}, 338 (2008)
  [arXiv:0801.2648 [hep-ph]];\\
  K.~Ichikawa, T.~Suyama, T.~Takahashi and M.~Yamaguchi,
  ``Non-Gaussianity, Spectral Index and Tensor Modes in Mixed Inflaton and
  Curvaton Models,''
  Phys.\ Rev.\  D {\bf 78}, 023513 (2008)
  [arXiv:0802.4138 [astro-ph]];\\
  T.~Suyama and F.~Takahashi,
  ``Non-Gaussianity from Symmetry,''
  JCAP {\bf 0809}, 007 (2008)
  [arXiv:0804.0425 [astro-ph]];\\
  X.~Gao,
  ``Primordial Non-Gaussianities of General Multiple Field Inflation,''
  JCAP {\bf 0806}, 029 (2008)
  [arXiv:0804.1055 [astro-ph]];\\
  M.~Beltran,
  ``Isocurvature, non-gaussianity and the curvaton model,''
  Phys.\ Rev.\  D {\bf 78}, 023530 (2008)
  [arXiv:0804.1097 [astro-ph]];\\
  B.~Chen, Y.~Wang and W.~Xue,
  ``Inflationary NonGaussianity from Thermal Fluctuations,''
  JCAP {\bf 0805}, 014 (2008)
  [arXiv:0712.2345 [hep-th]];\\
   M.~Li, T.~Wang and Y.~Wang,
  ``General Single Field Inflation with Large Positive Non-Gaussianity,''
  JCAP {\bf 0803}, 028 (2008)
  [arXiv:0801.0040 [astro-ph]];\\
  M.~Li, C.~Lin, T.~Wang and Y.~Wang,
  ``Non-Gaussianity, Isocurvature Perturbation, Gravitational Waves and a No-Go
  Theorem for Isocurvaton,''
  arXiv:0805.1299 [astro-ph];\\
  M.~Sasaki,
  ``Multi-brid inflation and non-Gaussianity,''
  Prog.\ Theor.\ Phys.\  {\bf 120}, 159 (2008)
  [arXiv:0805.0974 [astro-ph]];\\
  S.~Li, Y.~F.~Cai and Y.~S.~Piao,
  ``DBI-Curvaton,''
  arXiv:0806.2363 [hep-ph];\\
  Q.~G.~Huang,
  ``Spectral Index in Curvaton Scenario,''
  Phys.\ Rev.\  D {\bf 78}, 043515 (2008)
  [arXiv:0807.0050 [hep-th]];\\
  M.~Sasaki,
  ``Multi-brid inflation and non-Gaussianity,''
  Prog.\ Theor.\ Phys.\  {\bf 120}, 159 (2008)
  [arXiv:0805.0974 [astro-ph]];\\
  M.~Li and Y.~Wang,
  ``Consistency Relations for Non-Gaussianity,''
  JCAP {\bf 0809}, 018 (2008)
  [arXiv:0807.3058 [hep-th]];\\
  M.~Li and C.~Lin,
  ``Reconstruction of the isocurvaton scenario,''
  arXiv:0807.4352 [astro-ph];\\
  M.~Kawasaki, K.~Nakayama, T.~Sekiguchi, T.~Suyama and F.~Takahashi,
  ``Non-Gaussianity from isocurvature perturbations,''
  JCAP {\bf 0811}, 019 (2008)
  [arXiv:0808.0009 [astro-ph]];\\
  M.~Kawasaki, K.~Nakayama and F.~Takahashi,
  ``Non-Gaussianity from Baryon Asymmetry,''
  JCAP {\bf 0901}, 002 (2009)
  [arXiv:0809.2242 [hep-ph]];\\
  K.~Dimopoulos, D.~H.~Lyth and Y.~Rodriguez,
  ``Statistical anisotropy of the curvature perturbation from vector field
  perturbations,''
  arXiv:0809.1055 [astro-ph];\\
  T.~Moroi and T.~Takahashi,
  ``Non-Gaussianity and Baryonic Isocurvature Fluctuations in the Curvaton
  Scenario,''
  arXiv:0810.0189 [hep-ph];\\
  M.~Kawasaki, K.~Nakayama, T.~Sekiguchi, T.~Suyama and F.~Takahashi,
  ``A General Analysis of Non-Gaussianity from Isocurvature Perturbations,''
  arXiv:0810.0208 [astro-ph];\\
  T.~Matsuda,
  ``Non-standard kinetic term as a natural source of non-Gaussianity,''
  JHEP {\bf 0810}, 089 (2008)
  [arXiv:0810.3291 [hep-ph]];\\
  C.~T.~Byrnes, K.~Y.~Choi and L.~M.~H.~Hall,
  ``Large non-Gaussianity from two-component hybrid inflation,''
  arXiv:0812.0807 [astro-ph];\\
  C.~Hikage, K.~Koyama, T.~Matsubara, T.~Takahashi and M.~Yamaguchi,
  ``Limits on Isocurvature Perturbations from Non-Gaussianity in WMAP
  Temperature Anisotropy,''
  arXiv:0812.3500 [astro-ph];\\
  P.~D.~Meerburg, J.~P.~van der Schaar and P.~S.~Corasaniti,
  ``Signatures of Initial State Modifications on Bispectrum Statistics,''
  arXiv:0901.4044 [hep-th];\\
  H.~R.~S.~Cogollo, Y.~Rodriguez and C.~A.~Valenzuela-Toledo,
  ``On the Issue of the $\zeta$ Series Convergence and Loop Corrections in the
  Generation of Observable Primordial Non-Gaussianity in Slow-Roll Inflation.
  Part I: the Bispectrum,''
  JCAP {\bf 0808}, 029 (2008)
  [arXiv:0806.1546 [astro-ph]];\\
  Y.~Rodriguez and C.~A.~Valenzuela-Toledo,
  ``On the Issue of the $\zeta$ Series Convergence and Loop Corrections in the
  Generation of Observable Primordial Non-Gaussianity in Slow-Roll Inflation.
  Part II: the Trispectrum,''
  arXiv:0811.4092 [astro-ph].


\bibitem{Huang:2008rj}
  Q.~G.~Huang,
  ``The N-vaton,''
  JCAP {\bf 0809}, 017 (2008)
  [arXiv:0807.1567 [hep-th]].


\bibitem{Enqvist:2008gk}
  K.~Enqvist and T.~Takahashi,
  ``Signatures of Non-Gaussianity in the Curvaton Model,''
  JCAP {\bf 0809}, 012 (2008)
  [arXiv:0807.3069 [astro-ph]].


\bibitem{Huang:2008bg}
  Q.~G.~Huang and Y.~Wang,
  ``Curvaton Dynamics and the Non-Linearity Parameters in Curvaton Model,''
  JCAP {\bf 0809}, 025 (2008)
  [arXiv:0808.1168 [hep-th]].


\bibitem{Huang:2008zj}
  Q.~G.~Huang,
  ``Curvaton with Polynomial Potential,''
  JCAP {\bf 0811}, 005 (2008)
  [arXiv:0808.1793 [hep-th]].



\bibitem{Kawasaki:2008mc}
  M.~Kawasaki, K.~Nakayama and F.~Takahashi,
  ``Hilltop Non-Gaussianity,''
  arXiv:0810.1585 [hep-ph].


\bibitem{Langlois:2008vk}
  D.~Langlois, F.~Vernizzi and D.~Wands,
  ``Non-linear isocurvature perturbations and non-Gaussianities,''
  JCAP {\bf 0812}, 004 (2008)
  [arXiv:0809.4646 [astro-ph]].





\bibitem{Dimopoulos:2003az}
  K.~Dimopoulos, D.~H.~Lyth, A.~Notari and A.~Riotto,
  ``The curvaton as a Pseudo-Nambu-Goldstone boson,''
  JHEP {\bf 0307}, 053 (2003)
  [arXiv:hep-ph/0304050];\\
  E.~J.~Chun, K.~Dimopoulos and D.~Lyth,
  ``Curvaton and QCD axion in supersymmetric theories,''
  Phys.\ Rev.\  D {\bf 70}, 103510 (2004)
  [arXiv:hep-ph/0402059].

\bibitem{Svrcek:2006yi}
  P.~Svrcek and E.~Witten,
  ``Axions in string theory,''
  JHEP {\bf 0606}, 051 (2006)
  [arXiv:hep-th/0605206].





\bibitem{Starobinsky:1986fxa}
  A.~A.~Starobinsky,
  ``Multicomponent de Sitter (Inflationary) Stages and the Generation of
  Perturbations,''
  JETP Lett.\  {\bf 42} (1985) 152;\\
  M.~Sasaki and E.~D.~Stewart,
  ``A General Analytic Formula For The Spectral Index Of The Density
  Perturbations Produced During Inflation,''
  Prog.\ Theor.\ Phys.\  {\bf 95}, 71 (1996)
  [arXiv:astro-ph/9507001];\\
  M.~Sasaki and T.~Tanaka,
  ``Super-horizon scale dynamics of multi-scalar inflation,''
  Prog.\ Theor.\ Phys.\  {\bf 99}, 763 (1998)
  [arXiv:gr-qc/9801017];\\
    D.~H.~Lyth, K.~A.~Malik and M.~Sasaki,
  ``A general proof of the conservation of the curvature perturbation,''
  JCAP {\bf 0505}, 004 (2005)
  [arXiv:astro-ph/0411220];\\
  D.~H.~Lyth and Y.~Rodriguez,
  ``The inflationary prediction for primordial non-gaussianity,''
  Phys.\ Rev.\ Lett.\  {\bf 95}, 121302 (2005)
  [arXiv:astro-ph/0504045].




\bibitem{Sasaki:2006kq}
  M.~Sasaki, J.~Valiviita and D.~Wands,
  ``Non-gaussianity of the primordial perturbation in the curvaton model,''
  Phys.\ Rev.\  D {\bf 74}, 103003 (2006)
  [arXiv:astro-ph/0607627].



\bibitem{Bunch:1978yq}
  T.~S.~Bunch and P.~C.~W.~Davies,
  ``Quantum Field Theory In De Sitter Space: Renormalization By Point
  Splitting,''
  Proc.\ Roy.\ Soc.\ Lond.\  A {\bf 360} (1978) 117;\\
  A.~Vilenkin and L.~H.~Ford,
  ``Gravitational Effects Upon Cosmological Phase Transitions,''
  Phys.\ Rev.\  D {\bf 26}, 1231 (1982);\\
  A.~D.~Linde,
  ``Scalar Field Fluctuations In Expanding Universe And The New Inflationary
  Universe Scenario,''
  Phys.\ Lett.\  B {\bf 116}, 335 (1982);\\
  A.~A.~Starobinsky and J.~Yokoyama,
  ``Equilibrium state of a selfinteracting scalar field in the De Sitter
  background,''
  Phys.\ Rev.\  D {\bf 50}, 6357 (1994)
  [arXiv:astro-ph/9407016].

\bibitem{Suyama:2007bg}
  T.~Suyama and M.~Yamaguchi,
  ``Non-Gaussianity in the modulated reheating scenario,''
  Phys.\ Rev.\  D {\bf 77}, 023505 (2008)
  [arXiv:0709.2545 [astro-ph]].


\end{thebibliography}
\end{document}